\documentclass[amsmath,eqsecnum,preprint,pre,aps,showpacs]{revtex4-1}
\usepackage{amsmath}
\usepackage{graphicx}
\usepackage{epstopdf}
\usepackage{dcolumn}
\usepackage{bm}
\usepackage{float}

\usepackage{subfigure}
\usepackage[
    bookmarks=true,
    bookmarksopen=true,
    colorlinks=true,
    anchorcolor=blue,
    filecolor=blue,
    citecolor=blue,
    urlcolor=black,
    linkcolor=red,
    menucolor=blue,
    breaklinks=true
]{hyperref}

\begin{document}

\draft

\title{Structural relaxation, dynamical arrest \\ and aging in 
soft-sphere liquids}

\author{P. Mendoza-M\'endez$^{1}$, Ricardo Peredo-Ortiz$^{1,2}$,
Edilio L\'azaro L\'azaro$^{1,2}$, M. Ch\'avez-Pae$z^{2}$,
H. Ruiz-Estrada$^{1}$, F. Pacheco-V\'azquez$^{3}$, 
M. Medina-Noyola$^{2}$ and L.F. Elizondo-Aguilera$^{3 \ast}$}

\affiliation{$^{1}$ Facultad de Ciencias F\'isico-Matem\'aticas,
Benem\'erita Universidad Aut\'{o}noma de Puebla, Apartado Postal 
1152, CP 72000, Puebla, PUE, M\'{e}xico}

\affiliation{$^{2}$ Instituto de F\'{\i}sica,
Universidad Aut\'{o}noma de San Luis Potos\'{\i}, \'{A}lvaro
Obreg\'{o}n 64, 78000 San Luis Potos\'{\i}, SLP, M\'{e}xico}

\affiliation{$^3$ Instituto de F\'isica, Benem\'erita Universidad 
Aut\'onoma de Puebla, Apartado Postal J-48 72520, Puebla, M\'exico}
\email{ luisfer.elizondo@gmail.com}

\date{\today}

\begin{abstract}

We investigate the structural relaxation of a soft-sphere liquid 
quenched isochorically ($\phi=0.7$) and instantaneously to different 
temperatures $T_f$ above and below the glass transition.
For this, we combine extensive Brownian dynamics simulations and theoretical calculations based on the non-equilibrium self-consistent generalized Langevin equation (NE-SCGLE) theory. The response of the liquid 
to a quench generally consists of a sub-linear increase of the $\alpha$-relaxation time with system's age. 
Approaching the ideal glass-transition temperature from above 
($T_f>T^a$) sub-aging appears as a transient process describing 
a broad equilibration crossover for quenches to nearly arrested
states. This allows us to empirically determine an equilibration 
timescale $t^{eq}(T_f)$ that becomes increasingly longer as $T_f$ 
approaches $T^a$.
For quenches inside the glass ($T_f\leq T^a$) the growth rate of 
the structural relaxation time becomes progressively larger 
as $T_f$ decreases and, unlike the equilibration scenario, 
$\tau_{\alpha}$ remains evolving within the whole observation 
time-window.
These features are consistently found in theory and simulations 
with remarkable semi-quantitative agreement, and coincide with those 
revealed in the similar and complementary exercise [Phys. Rev. {\bf 96}, 
022608 (2017)] that considered a sequence of quenches with fixed final 
temperature $T_f=0$ but increasing $\phi$ towards the hard-sphere 
dynamical arrest volume fraction $\phi^a_{HS}=0.582$.
The NE-SCGLE analysis, however, unveils various fundamental aspects 
of the glass transition, involving the abrupt passage from the ordinary 
equilibration scenario to the persistent aging effects that are 
characteristic of glass-forming liquids. The theory also explains 
that, within the time window of any experimental observation, this 
can only be observed as a continuous crossover.

\end{abstract}


\maketitle

\section{Introduction.}\label{sectionI}

Aging is a fundamental aspect of the process of amorphous solidification
observed in glass and gel-forming liquids upon undercooling
\cite{cipelletti1,cipelletti2,lunkenheimer,di,bandyopadhyay,bissing,
hecksher}.
It is embodied in a slow and long-lasting evolution of the physical
observables with the waiting time $t$ after preparation, and it is
consistently found in a wide variety of microscopically very different
materials, ranging from polymers \cite{hutchinson,bellon}, metallic alloys
\cite{das,ruta}, aqueous clay suspensions \cite{knaebel,ruzicka,angelini} 
and colloids \cite{cipelletti3,jain,gordon,jacob,bonn,abou}, thus 
revealing its universal character.

In contrast with the number of experimental
\cite{elmasriaging,martinezvanmegen,sanz1,lu1} and computer simulation
\cite{kobbarrat,foffiaging1,puertas1,warren} investigations carried out 
to quantify the irreversible relaxation in a variety of model systems 
during aging, only a few first-principles theoretical approaches 
\cite{degregorio,cugliandolo1} are available for the description and 
understanding of such out-of-equilibrium processes.
In particular, about two decades ago Latz \cite{latz1,latz2,latz3} 
formally extended the well established mode coupling theory (MCT) of 
the glass transition (GT) \cite{goetze1,goetze2,goetze3,goetze4} to 
allow for the description of aging. However, no quantitative predictions 
have been presented so far that could be contrasted with experimental 
or simulation results for specific model systems of structural 
glass-formers. Thus, one can state that until recently, no quantitative 
first-principles theory has been developed and applied to successfully 
describe aging phenomena in structural glass-forming (atomic or colloidal) 
liquids.

In this regard, similar developments to those made by Latz were
carried out in the context of the so-called self-consistent generalized
Langevin equation (SCGLE) theory of colloid dynamics
\cite{scgle1,scgle2,scgle3,scgle4,scgle5} and dynamical arrest
\cite{arrest1,arrest2,arrest3}. In this manner, this theory, which can be regarded
as an alternative to MCT \cite{thvg_elizondo}, could be extended 
to non-equilibrium conditions, allowing the description of the spatially 
non-uniform and temporally non-stationary relaxation of glass-forming 
liquids. Such an extension, carried out in Ref. \cite{nescgle1} and 
referred to as the \emph{non-equilibrium} self-consistent generalized 
Langevin equation (NE-SCGLE) theory, led to the emergence of a 
quantitative first-principles description of aging in structural 
glass formers.

The theoretical framework provided by the NE-SCGLE has solidly
demonstrated its ability to predict some of the most relevant 
universal signatures of both, the glass and gel transitions, 
including aging effects, as well as very specific features 
reflecting the particular role of molecular interactions in 
the explicit systems considered.
For example, this theory accurately describes the process of 
formation of high-density hard-sphere--like glasses
\cite{nescgle2,nescgle3,gabriel,nescgle5},
whereas for low-density and low-temperature liquids with excluded volume  
plus attractive interactions it predicts the formation of sponge-like 
gels and porous glasses by arrested spinodal decomposition
\cite{olais1,olais2,olais3,zepeda}. The comparison of such predictions
with both, experimental and simulated data for systems with essentially
the same kind of molecular interactions, confirms that the NE-SCGLE
is a successful non-equilibrium statistical mechanics approach for 
the description of the slowing down and the aging of the dynamics 
observed in glass- and gel-forming liquids.

Let us refer, in particular, to the scenario predicted by this theory
when the Weeks-Chandler-Andersen (WCA) soft-sphere liquid  is suddenly 
and isochorically quenched to zero temperature within the constraint 
that it remains spatially uniform. This scenario was first systematically 
analyzed using the NE-SCGLE theory in Ref. \cite{nescgle3}. Such study 
considers a series of independent quench processes that differ only in 
the value of the fixed density $n$.
Each of these processes starts with the repulsive liquid in an equilibrium
(fluid-like) state of density $n$ and initial temperature $T_{i}$.
Afterwards, the system is instantaneously quenched to a final temperature
value $T_{f}=0$. Since the zero-temperature limit of such soft-sphere 
fluid is actually a hard-sphere (HS) liquid of volume fraction $\phi$, 
these processes mimic the spontaneous search for the equilibrium state 
of a HS liquid, driven to non-equilibrium conditions by some perturbation
(shear, for example \cite{brambilla,elmasri}) that ceases at a time $t=0$.

For each of these isochoric processes, the NE-SCGLE described
the irreversible relaxation of the system in terms of the 
time-evolution of the \emph{non-equilibrium} structure factor (SF) 
$S(k;t)$ and of dynamical properties, such as the \emph{self} 
intermediate scattering function (ISF) $F_S(k,\tau;t)$, where $\tau$ 
is the correlation (or \emph{delay}) time and $t$ is the waiting 
(or \emph{evolution}) time after the quench. 
In particular, the predicted $t$-development of the $\alpha$-relaxation 
time $\tau_\alpha(t)$ (with $F_S(k,\tau_\alpha;t)=1/e$) 
allowed for the identification of an \emph{equilibration time} 
$t^{eq}(\phi)$, defined as the time-scale after which $\tau_\alpha(t)$ 
saturates to its equilibrium value $\tau_{\alpha}^{eq}(\phi)$.
Thus, according to the NE-SCGLE one possibility is that the quenched 
hard-sphere liquid will recover equilibrium within a finite time window 
$t^{eq}(\phi)$ that depends on the specific value of the (fixed) volume 
fraction $\phi$.
However, both $t^{eq}(\phi)$ and $\tau_{\alpha}^{eq}(\phi)$ are also 
predicted to diverge as $\phi$ approaches the HS dynamical-arrest 
volume fraction $\phi^{a}_{HS}$ from below. Let us recall that, 
according to the \emph{equilibrium} SCGLE theory \cite{gabriel}, the 
HS system remains ergodic for $0< \phi <\phi^{a}_{HS} = 0.582$ and 
becomes dynamically arrested at $\phi = \phi^{a}_{HS}$ and beyond. 
Hence, for quenches with $\phi\geq\phi^{a}_{HS}$, the prediction is  
that the system will no longer be able to equilibrate, but instead, 
will age forever during the endless process of becoming a glass.

Furthermore, another relevant prediction of the NE-SCGLE theory for the
quench processes considered is that, for a fixed \emph{finite} waiting 
time $t$, the plot of $\tau_\alpha(t;\phi)$ as a function of $\phi$ 
exhibits two regimes, corresponding to samples that have fully 
equilibrated within this waiting time $(\phi\le \phi_{0}(t))$, and to 
samples for which equilibration is not yet complete 
$(\phi\ge \phi_{0}(t))$. The crossover volume fraction $\phi_{0}(t)$ 
increases with $t$ but saturates to the value $\phi^{a}_{HS}$. These 
predictions have been contrasted against analogous results of extensive 
molecular dynamics simulations finding a remarkable consistency 
\cite{gabriel,nescgle5}.

The work in Refs. \cite{gabriel,nescgle3,nescgle5}, however, leaves 
open several important issues. One of them refers to the fact that 
the physical scenario just described is conceptually closer to the 
spontaneous equilibration or aging of a non-equilibrium HS colloidal 
dispersion \cite{brambilla,elmasri}, in which the volume fraction 
$\phi$ is the relevant control parameter.
However, in other aging experiments involving actual soft-sphere 
colloidal suspensions (made of, \emph{e.g.,} thermosensitive microgel 
colloids \cite{sessoms,rivas}) it is temperature (and/or softness) 
rather than volume fraction, the relevant control parameter. Such 
colloidal quench experiments are, in fact, conceptually closer to the 
equilibration and aging of atomic glass-formers \cite{di}. Experience
indicates that in thermally-driven systems the inverse temperature 
$T^{-1}$ plays a qualitatively analogous role to that of the volume 
fraction $\phi$ in density-driven colloidal systems. Thus, an obvious 
question is whether the NE-SCGLE theory has the ability to explain in 
more precise terms this conceptually important empirical analogy. Addressing this well-defined challenge constitutes the main objective of the present contribution.

With this objective in mind, we have carried out extensive Brownian 
dynamics simulations, as well as theoretical calculations within the 
NE-SCGLE framework, which we combine to study the irreversible 
relaxation and the GT in the same soft-sphere system as in Refs. 
\cite{gabriel,nescgle3,nescgle5}. As just mentioned, these references 
involved a sequence of isochoric quenches to the same final temperature 
$T_f=0$, but along different isochores (\emph{i.e.,} different volume 
fractions $\phi$). In contrast, in this work we investigate the 
complementary study, involving a sequence of isochoric quenches, now along a given 
isochore with fixed volume fraction $\phi\ (>\phi^{a}_{HS} = 0.582)$, 
but with different final $T_f$, both above and below the dynamical 
arrest line $T=T^a(\phi)$ (solid curve in Fig. \ref{fig:1} below). 
We then analyze the concomitant relaxation of the system in terms of 
the waiting-time evolution of the structural and dynamical properties provided by both, simulations and theory, as a function of the control parameter $T_f$.

As in Ref. \cite{nescgle1}, here we find two qualitatively different
scenarios for the relaxation of the dynamics, identified through the
specific evolution of the $\alpha$-relaxation time $\tau_{\alpha}(t)$ after quenching.
The first involves quenches with $T_f>T^a(\phi)$, for which 
$\tau_{\alpha}(t)$ shows a transient increase with system's age before 
attaining a new stationary equilibrium value $\tau_{\alpha}^{eq}(T_f)$.
This behavior, in turn, allows us to obtain an empirical relation 
for the equilibration time $t^{eq}(T_f)$ of the system, which grows as 
a power law of $\tau_{\alpha}^{eq}(T_f)$ with exponent larger than one.
A second scenario is found for quenches with $T_f\leq T_a(\phi)$, in 
which we observe that $\tau_{\alpha}(t)$ grows sub-linearly with 
waiting time and, unlike the equilibration scenario, the structural
relaxation time exhibits a persistent development over the entire 
observation time-window, showing also an increasingly larger growth 
rate as $T_f$ becomes smaller. All these features are consistently 
observed in simulations and theory, finding only small deviations at 
the longest waiting times explored. 

The main results of this contribution are conveniently summarized in 
a plot of $\tau_\alpha(t;1/T)$ vs $1/T$, thus allowing us to compare 
the overall physical scenario obtained here with that previously 
reported in Refs. \cite{nescgle5,gabriel}, corresponding to the 
hard-sphere limit ($T_f=0$) of the same glass-forming liquid. Such 
a comparison elucidates the qualitative correspondence of the two 
control parameters $\phi$ and $1/T$ in determining the non-equilibrium 
relaxation of a soft-sphere liquid quenched towards the GT.  

Our work is organized as follows: 
In section \ref{SEC1} we briefly describe the model soft-sphere liquid 
studied with simulations and theory. To serve as a reference, in the 
same section we also present the arrested states diagram predicted by 
the NE-SCGLE for such system, along with the sequence of quenches 
employed to investigate its irreversible relaxation. 
In section \ref{SEC2} we focus on the description of the process of 
equilibration at the level of both structural and dynamical properties, 
with the former allowing us to determine the aforementioned equilibration 
time-scale of the system. 
Section \ref{SEC3} addresses the description of the crossover from 
equilibration to the dynamical arrest and aging regime. For clarity and 
methodological convenience, this section is divided in three parts. In 
subsection \ref{agingBD} we present and discuss the main results of our 
BD simulations for the kinetic evolution of the system for quenches below 
the critical temperature $T^a$. 
In subsections \ref{agingSC1} and \ref{agingSC2}, respectively, the 
non-equilibrium relaxation of the structural and dynamical properties 
during the development of glassy states is explained and discussed within 
the theoretical framework provided by the NE-SCGLE. In section \ref{SEC4} 
we describe the connection between our results and previous findings for 
the HS limit of the same model system considered here, thus allowing to 
establish the qualitative correspondence of the two control parameters 
$\phi$ and $T$ in determining the non-equilibrium relaxation of 
the quenched soft-sphere fluid. 
Finally, in Sec. \ref{conclusions} we summarize our main conclusions.
For clarity, a brief review of the NE-SCGLE equations is provided in
appendix \ref{appendix_theo}, whereas the main details of our BD 
simulations are described in appendix \ref{sim_details}.

\section{Temperature-quenched soft-sphere system: dynamical arrest diagram}\label{SEC1}

In in this work we will study the structural relaxation of a 
monocomponent Brownian fluid constituted by $N$ soft-spheres of 
diameter $\sigma$ in a volume $V$, interacting through a 
Weeks-Chandler-Anderson (WCA) potential \cite{wca}, that vanishes 
for $r$ larger than the distance $\sigma$ of soft contact, but which 
for $r\leq\sigma$ is given by

\begin{equation}
 u(r) =
\epsilon\left[ \left( \frac{\sigma}{r}\right)^{12}
-2\left( \frac{\sigma}{r}\right)^{6}+1 \right].
\label{truncatedlj}
\end{equation}

\noindent The state space of this model system is spanned by the volume
fraction $\phi = \pi n \sigma^3/6$ (with $n=N/V$) and the effective
temperature $T^*\equiv k_BT /\epsilon$. For simplicity, from now on
we shall refer to the dimensionless parameter $T^*$ just as $T$ and
use $\sigma$ and [$\sigma^2/D^0$] as the units of length and time,
respectively, with $D^0$ being the short-time self-diffusion coefficient,
related to the corresponding short-time friction coefficient $\zeta^0$ by 
Einstein's relation $\zeta^0=k_BT/D^0$. For colloidal liquids $\zeta^0$ can 
be determined by its conventional Stokes expression, whereas for atomic 
liquids $\zeta^0$ represents the less intuitive Doppler friction 
\cite{paty,ornsteinuhlenbeck}.

\begin{figure}
\includegraphics[scale=.4]{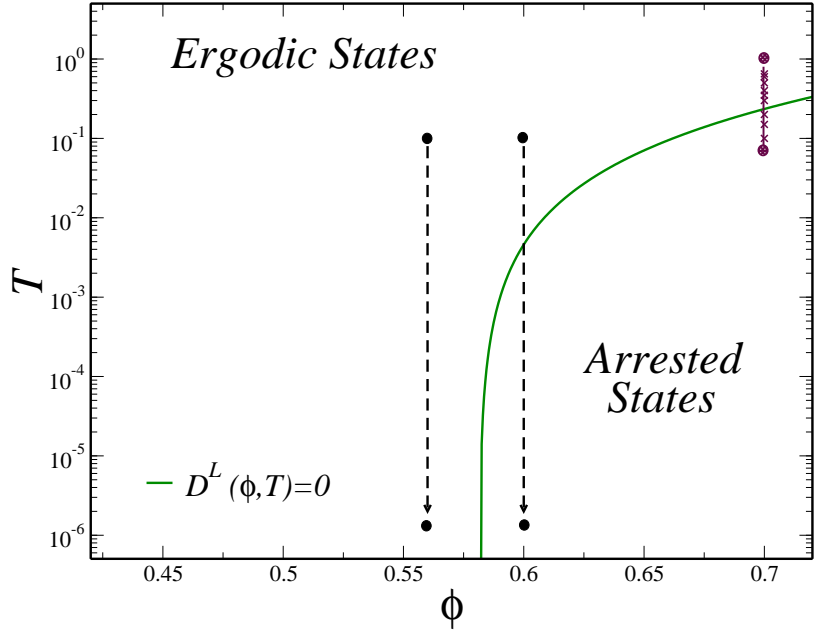}
\caption{Dynamical arrest diagram of the WCA model defined in Eq.
\eqref{truncatedlj}, predicted by the NE-SCGLE using the numerical
solution of the Ornstein-Zernike equation and the 
Percus-Yevick/Verlet-Weis approximation \cite{percus,verlet} for
the determination of the thermodynamical inputs required by the theory
(see Sec. IIIA of Ref. \cite{nescgle3}). The solid line that separates 
the regions of ergodic and arrested states corresponds to the locus of 
the ideal isodiffusivity line $D^L(\phi,T)=0$.
The dashed vertical arrows on the left represent the two qualitatively 
distinct types of quench processes studied in Ref \cite{nescgle3}, 
corresponding to isochoric quenches at fixed $\phi=0.56<\phi_{HS}^a$ 
and $\phi=0.6>\phi_{HS}^a$, from an initial temperature $T_i=0.1$ and 
with final value $T_f=0$, thus describing the processes of equilibration 
and aging in a HS fluid, respectively. The solid arrow in the top right 
represents, instead, a sequence of isochoric quenches at fixed 
$\phi=0.7$, all of which start from the same initial temperature 
$T_i=1$, but end at different values of $T_f$, used here to describe 
the crossover from equilibration, 
$T_f>T^a(\phi=0.7)$, to aging, $T_f\leq T^a(\phi=0.7)$, with 
$T^a(\phi=0.7)=0.235$.}
\label{fig:1}\end{figure}

To organize our discussion, let us start by considering Fig. \ref{fig:1},
which displays the dynamical arrest diagram predicted by the NE-SCGLE 
theory for the WCA fluid just described. Except for trivial modifications 
explained below, this figure is Fig. 1 of  Ref. \cite{nescgle3}. 
The same reference provides all the pertinent numerical details. 
This diagram identifies two different regions of the ($\phi,T$)-plane, 
corresponding to asymptotically stationary ergodic (fluid) and 
non-ergodic (dynamically arrested) states. The boundary between these 
regions is delimited by the ideal isodiffusivity line $D^L(\phi,T)=0$, 
with $D^L$ being the long-time diffusion coefficient. 
This condition defines a dynamical arrest temperature $T^a(\phi)$ which, 
for $\phi>\phi_{HS}^a$, describes a monotonically increasing function of 
the packing fraction (solid curve in Fig. \ref{fig:1}).
As we show in what follows, for any isochoric quench from the ergodic
region (with $\phi>\phi_{HS}^a$) the critical value $T^a(\phi)$ separates
two possible and mutually exclusive regimes for the relaxation of the 
dynamics of the WCA system, corresponding to equilibration (for quenches 
with $T_f>T^a(\phi)$) or dynamical arrest and aging (for 
$T_f\leq T^a(\phi)$), with $T^a(\phi=0.7)=0.235$.

As already mentioned, these two scenarios were analyzed in detail in
Ref. \cite{nescgle3}, but only for the specific case in which the system
defined in Eq.\eqref{truncatedlj} is instantaneously quenched from the 
ergodic region to a final temperature $T_f=0$. 
As emphasized in the introduction, in this work we are interested in the
study of the complementary category of thermally-driven processes that
consider, instead, quenches to a finite value $T_f\neq0$. Thus, in what
follows we analyze a sequence of isochoric quenches ($\phi=0.7$), all of
which start from the same state point in the high density/high temperature
ergodic regime and whose final temperature $T_f$ gradually approaches and 
crosses the critical value $T^a(\phi=0.7)$. These processes are 
illustrated by the solid arrow and cross symbols in the top right of 
Fig.\ref{fig:1}.

\section{Equilibration of the WCA model after a quench.}\label{SEC2}

For clarity and methodological convenience, we will first discuss the
conceptually simplest scenario for the relaxation of the WCA liquid.
We refer to the process of full equilibration, corresponding to the case
in which this system, originally at equilibrium, is quenched to a final
state that also lies in the ergodic region (\emph{i.e.,} $T_f>T^a(\phi)$).
As discussed in Refs. \cite{nescgle1,nescgle2,nescgle3,gabriel}, the
relaxation of the quenched liquid can be described in terms of the
time-evolution of the non-equilibrium SF, $S(k;t)\equiv\langle\delta 
n(\mathbf{k};t)\delta n^{\dagger}(-\mathbf{k};t)\rangle$, with
$\delta n(\mathbf{k};t)$ being the Fourier transform of the fluctuations
in the number density
$n(\mathbf{r};t)=\sum_{i=1}^N\delta(\mathbf{r}-\mathbf{r}_i(t))/\sqrt{N}$,
and also in terms of the \emph{self} intermediate scattering function (ISF)
$F_S(k;\tau;t)=\langle 
\exp[i\mathbf{k}\cdot\Delta\mathbf{r}_T(\tau,t)]\rangle$, with
$\Delta\mathbf{r}_T(\tau,t)\equiv[\mathbf{r}_T(t+\tau)-\mathbf{r}_T(t)]$ 
being the displacement of a tracer particle and where the angular brackets 
indicate average over a time-dependent non-equilibrium statistical 
ensemble. In what 
follows, we will stick to the analysis of these two properties. 
For reference, a brief summary of the NE-SCGLE equations describing the 
$t$-evolution of these quantities is provided in Appendix 
\ref{appendix_theo}. The details of our BD simulations are provided in 
Appendix \ref{sim_details}.

Before discussing our results, let us recall that solving the NE-SCGLE 
equations (Eqs. \eqref{relsigmadif2pp}-\eqref{lambdadk} of Appendix  
\ref{appendix_theo}) requires the previous determination of the fundamental 
thermodynamic input $ \mathcal{E}(k;n,T)$ involved in Eq.  
\eqref{relsigmadif2pp}, related to the FT of the two-particle direct 
correlation function $c[|\mathbf{r}-\mathbf{r}'|;n;T]$, and to the  
equilibrium structure factor $S^{eq}(k;n,T)$ by 
$n\mathcal{E}(k;n,T)=1-n c(k;n,T)=1/S^{eq}(k;n,T)$.  As explained in 
detail in Ref. \cite{nescgle3}, in our present case this property can 
be determined using the Ornstein-Zernike (OZ) equation combined with 
the Percus-Yevick/Verlet-Weiss (PY-VW) approximation \cite{percus,verlet} 
for an effective HS fluid with a $T$-depedent HS diameter $\sigma(T)$ 
obtained by the blip function method \cite{hansen} (see also Sec. IIIA 
of Ref. \cite{nescgle3}). The demonstrative diagram of Fig. \ref{fig:1} 
was in fact calculated in this manner, leading to the value 
$T^a(\phi=0.7)=0.235$ of the dynamic arrest temperature along the isochore 
$\phi=0.7$ quoted above. In what follows, however, we shall introduce a 
slight modification of this procedure  to fine-tune the quantitative 
comparison between theory and simulations. 

Thus, to analyze relaxation processes corresponding to comparable 
initial and final states, we have fine-tuned the previous approximate 
determination of the thermodynamic input, which we denote as 
$\mathcal{E}_{PY/VW}(k;n,T)$. For this we used well-equilibrated  
simulation results for the static structure factor $S_{BD}^{eq}(k;
\phi=0.7,T)$ at a set of values of the temperature $T$, to determine a 
rescaled theoretical temperature  $T'(T)$ such that the height of the main 
peak of the theoretical structure factor $S_{PY/VW}^{eq}(k;\phi=0.7,T') 
\equiv 1/ n\mathcal{E}_{PY/VW}(k;\phi=0.7,T')$ fits the height of 
the main peak of  $S_{BD}^{eq}(k;\phi=0.7,T)$. This procedure leads to the 
approximate rescaled temperature  $T'=(3.2)T^{0.785}$. In addition, 
following previous work \cite{nescgle5,gabriel} we have also employed the 
simulation values of the equilibrium mobility $b^{eq}(\phi=0.7,T)$ to 
calibrate  the only free parameter of the NE-SCGLE, namely, the cutoff 
wave-vector $k_c$  (see Eq. \eqref{lambdadk}), leading to the choice 
$k_c=2.05 \times 2\pi/\sigma$. The changes resulting from the previous 
rescaling of $T$ and calibration of $k_c$ are only quantitative. For 
example, the new value of the dynamic arrest temperature along the isochore 
$\phi=0.7$ is  $T^a(\phi=0.7)=0.166$ (compared to the previous value 
$T^a(\phi=0.7)=0.235$ quoted above).

\begin{figure*}[h!]
\includegraphics[width=0.3\textwidth, height=0.3\textwidth]{agrs/new_fig2a.eps}
\includegraphics[width=0.3\textwidth, height=0.3\textwidth]{agrs/new_fig2b.eps}
\includegraphics[width=0.3\textwidth, height=0.3\textwidth]{agrs/new_fig2c.eps}\\
\includegraphics[width=0.3\textwidth, height=0.3\textwidth]{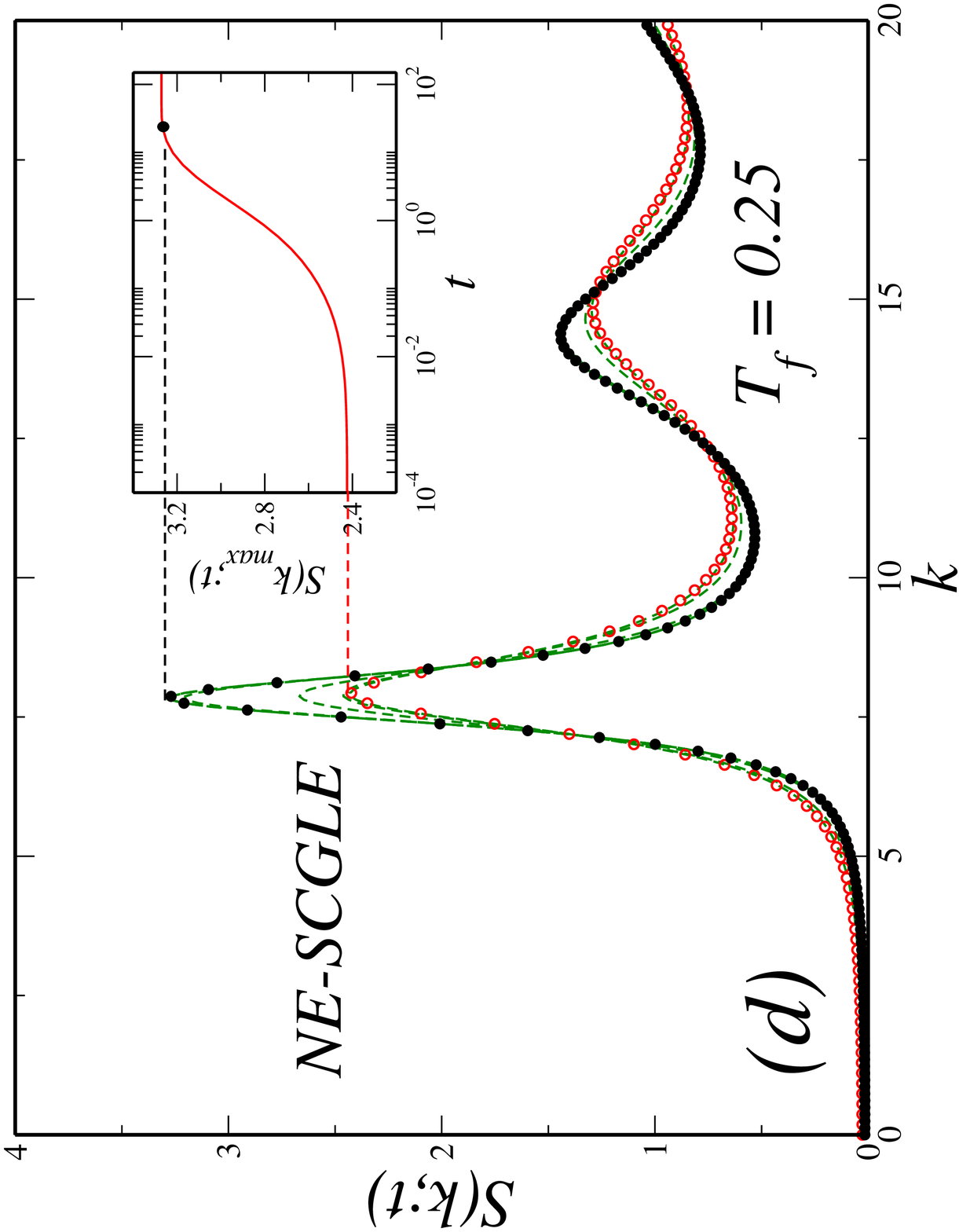}
\includegraphics[width=0.3\textwidth, height=0.3\textwidth]{agrs/Final_Fig2E.eps}
\includegraphics[width=0.3\textwidth, height=0.3\textwidth]{agrs/Final_Fig2F.eps}
  \caption{Schematic representation of an equilibration process of 
  the WCA fluid after an instantaneous isochoric quench from the 
  equilibrium state point ($\phi=0.7,T_i=1$) to a final temperature 
  $T_f$ above the dynamical arrest point $T^a=0.166$. (a) and 
  (d) Sequence of snapshots at waiting times $t=10^{-3},10^{-2},10^{-1},
  10^0,10^1$ and $10^2$ (green dashed lines) obtained from BD 
  simulations and theory, respectively, describing the waiting time
  evolution of the SF $S(k,t)$ from its initial value,
  $S_i(k)=S(k;t=0)=S^{eq}(k;\phi=0.7,T_i=1)$ (open symbols), towards 
  a final equilibrium value $S_f(k)=S^{eq}(k;\phi=0.7,T_f=0.25)$ 
  (solid symbols).
  In both figures, the inset emphasizes the time-evolution of the 
  maximum $S_{max}(t)\equiv S(k=k_{max};t)$ of the SF. (b) and (e)
  Corresponding sequence of snapshots for the self-ISF $F(k,t)$. 
  In both figures, the inset highlights the evolution of the 
  $\alpha$-relaxation time $\tau_{\alpha}(t)$, scaled as 
  $\tau^*_{\alpha}(t)\equiv  k_{max}^2D^0\tau_{\alpha}(t)$, for the specific 
  quench considered. (c) and (f) Waiting time dependence of  
  $\tau^*_{\alpha}(t)$ for a sequence of quenches at various 
  $T_f$ above and approaching the transition temperature $T^a(\phi=0.7)$.
  The dark circles along the dashed lines in both figures illustrate 
  the relationship between the equilibration time 
  $t^{eq}(\phi=0.7,T_f)$ and the equilibrium $\alpha$-relaxation time 
  $\tau^{eq}_{\alpha}(\phi=0.7,T_f)$ of each quench (see text).}
  \label{fig:2}
\end{figure*}

Fig. \ref{fig:2} summarizes the results obtained with both, simulations 
(upper panels) and theory (lower panels), for the description of the 
process of equilibration of the WCA fluid after an instantaneous 
isochoric quench from the initial state point ($\phi=0.7,T_i=1.0$) to 
a final temperature $T_f=0.25$.
Figs. \ref{fig:2}(a) and \ref{fig:2}(d), for instance, display a sequence
of snapshots at waiting times $t=0,10^{-3},10^{-2},10^{-1},10^0,10^1$ and
$10^2$ describing the evolution of the SF after quenching.
At $t=0$ (open symbols in both figures) the function $S_i(k)$ shows the 
typical behavior of a dense repulsive liquid, mainly characterized by the 
development of a well defined peak at $k = k_{max}\approx7.7$, and
oscillations around the unitary value for larger wave-vectors.
After the quench ($t>0$), the SF gradually evolves describing an ordinary 
de Gennes narrowing \cite{degennes}, which consists of a notorious 
amplification of the structural correlations around $k_{max}$ (see the 
inset of both figures) and also a small shift of the second peak towards 
slightly smaller $k$ values.
At a time-scale of order $10^1$, $S(k,t)$ ceases to evolve and reaches 
a new (equilibrium) stationary value $S_f(k)$ (solid symbols).
All these features, which describe the typical structural behavior of 
a soft-sphere fluid upon cooling, are consistently found in both,
simulations and theory.

As shown in Figs. \ref{fig:2}(b) and \ref{fig:2}(e), for this kinetic 
process one also finds a gradual slowing down in the dynamics. This 
is observed in the increasingly slower $\tau$-relaxation displayed by 
the ISF $F_S(k,\tau;t)$ as $t$ becomes larger or, equivalently, in the 
development of the $\alpha$-relaxation time $\tau_\alpha(k;t)$ 
(defined by the condition $F_S(k,\tau_{\alpha};t)=1/e$) with waiting-time.
One notices, for instance, that $\tau_\alpha(t)$ grows roughly 
one order of magnitude during the time-scale of evolution of the 
SF (see the insets of Figs. \ref{fig:2}(b) and \ref{fig:2}(e)). 
For larger $t$, the ISF stops evolving and reaches a new stationary 
value $F_S^{eq}(k;\tau)$ (solid symbols), so that the $\alpha$-relaxation 
time saturates to a constant $\tau_{\alpha}^{eq}(T_f)$, thus 
indicating the full equilibration of the system. 

In order to emphasize the role of the final temperature of a quench 
on the slowing down and equilibration of the dynamics, we might 
consider the behavior of $\tau_{\alpha}(t)$ for a sequence of quenches 
with decreasing $T_f$.
This is done in Figs. \ref{fig:2}(c) and \ref{fig:2}(f), which show 
the results obtained with simulations and theory, respectively, 
for the evolution of the $\alpha$-relaxation time, scaled as 
$\tau^*_{\alpha}(t)\equiv k_{max}^2D^0\tau_{\alpha}(t)$ and for different 
values of $T_f$, as indicated.
Thus presented, these results highlight the transient increase in
the $\alpha$-relaxation time during times shorter than the 
$T_f$-dependent equilibration time $t^{eq}(T_f)$, after which 
$\tau_{\alpha}(t)$ has saturated at its corresponding equilibrium 
value $\tau^{eq}_{\alpha}(\phi=0.7,T_f)$. 
Notice that both, $\tau^{eq}_{\alpha}(\phi=0.7,T_f)$ and $t^{eq}(T_f)$ 
become increasingly larger with decreasing $T_f$. For reference, the 
equilibration times found in simulations and theory are represented 
in Figs. \ref{fig:2}(c) and \ref{fig:2}(f) by the solid black circles. 
One notices that these highlighted values align themselves to a good 
approximation along the dashed straight lines, which thus define an 
empirical relationship of the form 

\begin{equation}
\tau_{\alpha}^{eq}(T_f)\approx A\times[t^{eq}(T_f)]^{\gamma},
\label{empreltauvsteq}
\end{equation}

\noindent that can also be written as $t^{eq}(T_f)\propto 
[\tau_{\alpha}^{eq}(T_f)]^\eta$, with $\eta=1/\gamma$. From the 
Brownian dynamics simulation results in Fig. \ref{fig:2}(c) we 
determine the values $A_{BD}=0.13$ and $\gamma_{BD}=0.8063$, i.e., 
$t^{eq}\propto [\tau_{\alpha}^{eq}]^{1.24}$. These values may be 
compared with those determined in Ref. \cite{gabriel} from the 
molecular dynamics simulation of a sequence of quenches that 
approach dynamic arrest by increasing the volume fraction: 
$A_{MD}=2.5$ and $\gamma_{MD}=0.7$, which lead to $t^{eq}(T_f)\propto 
[\tau_{\alpha}^{eq}(T_f)]^{1.43}$. This means that these simulations 
(in agreement with others \cite{kimsaito}) establish that as the glass 
transition is approached, i.e., as  $t^{eq}$ and  $\tau_{\alpha}^{eq}$ 
increase without bound, the former will always increase faster than the 
latter.
Let us also highlight the fact that the apparent instantaneous slope 
$\mu(t)\equiv [d \log \tau_{\alpha} (t)/d\log t]$ of the simulation 
results for $\tau_{\alpha}(t)$ in Fig. \ref{fig:2}(c) is always smaller 
than the slope of the dashed straight line of the figure, which is 
itself smaller than unity. In other words, the simulated equilibration 
processes in Fig. \ref{fig:2}(c) indicate that $\mu(t) < 1.$ In 
addition, let us also notice that the simulation data for 
$ \log\tau_{\alpha}(t)$ as a function of $\log t$ exhibits a 
point of inflection, below which $\mu(t)$ increases with $t$, while 
above which it decreases. As $T_f$ becomes smaller, however, this 
inflection point evolves into a finite and increasingly larger 
$t$-interval in which $\mu(t)$ appears nearly constant, thus 
suggesting that $\tau_{\alpha}(t)$ grows as a power law of the form 
$\tau_{\alpha}(t)\propto t^\delta$ ($\delta<1$) within such interval. 
We will come back to this point later in Sec. \ref{compa}

In a first attempt to see if the NE-SCGLE theory captures these important 
conclusions, we first analyzed the data for $\tau_{\alpha}(t)$ in Fig. 
\ref{fig:2}(f) using the theoretical definition of the equilibration time 
$t^{eq}$ given in Eq. (4.5) of Ref. \cite{nescgle3}. As a result, we found 
that the theoretical predictions actually exhibited the opposite trend, 
namely, that  $t^{eq}\propto [\tau_{\alpha}^{eq}]^\eta$, with the exponent 
$\eta=1/\gamma=1/1.07=0.93$ being smaller, not greater, that one. Let us 
mention that this result is consistent with the NE-SCGLE result  $\eta=1/
\gamma=1/1.05=0.95$ determined in Ref.  \cite{nescgle3} for the quench 
processes studied there.  
Hence, both predictions oppose the trends established by simulations.
As it happens, however, such tentative NE-SCGLE results actually reflect 
the tentative definition of $t^{eq}$ proposed in Ref.  \cite{nescgle3}, 
rather than a poor accuracy of the NE-SCGLE predictions for 
$\tau_{\alpha}(t)$, as we confirmed by exploring other definitions of 
$t^{eq}$. 

Thus, inspired by reference \cite{warren} we finally adopted the use 
of the apparent instantaneous slope $\mu(t)$ to determine $t^{eq}$. 
Specifically, in terms of this quantity the equilibration condition 
is defined as $\mu(t^{eq};T_f)=0.1$. 
Analyzing the NE-SCGLE data for $\tau_{\alpha}(t)$ in Fig. 
\ref{fig:2}(f) with this new criterion, we determined the solid black 
circles in the figure, which align themselves according to the 
empirical fit 
$\tau_{\alpha}^{eq}(T_f)\approx 0.189 \times[t^{eq}(T_f)]^{0.795}$, which 
implies that $t^{eq}(T_f)\propto [\tau_{\alpha}^{eq}(T_f)]^{1.26}$, now 
in full agreement with simulations. For completeness, we also reanalyzed 
the quench processes studied in Ref.  \cite{nescgle3}, now applying this 
new definition of $t^{eq}$, with the result $\eta=1/\gamma=1/0.92=1.08$. 

Regarding other features of the simulations that are consistently 
described by the theory, let us stress that the apparent instantaneous 
slope $\mu(t)$ of the latter is also smaller than the slope of the 
dashed straight line of the figure, and that the apparent instantaneous 
slope $\mu(t)$ first increases with $t$, and beyond an inflection point, 
it decreases. Furthermore, as in the simulations, for deeper quenches, 
this inflection point also becomes a finite interval in which  
$\tau_{\alpha} (t)$ increases as a power law of $t$, $\tau_{\alpha}(t) 
\propto t^\delta$, with $\delta<1$. Thus, in conclusion, theory and 
simulations agree in the general physical scenario for the equilibration processes of the instantaneously quenched WCA liquid.

\section{Aging of a quenched soft-sphere system. }\label{SEC3}

We now discuss the other possible scenario for the irreversible 
relaxation of the WCA model, in which the system starts at the 
same initial equilibrium state as before, but is now quenched 
into the non-ergodic region (i.e., $T_f\leq T^a(\phi=0.7)$) 
where it is predicted to become dynamically arrested.
In Fig. \ref{fig:3} we summarize the results obtained with BD 
simulations (upper panels) and the NE-SCGLE (lower panels) for 
three representative quenches at the indicated temperatures.

\subsection{Aging of the structure and dynamics of the WCA model: 
Brownian dynamics.}\label{agingBD}

\begin{figure*}
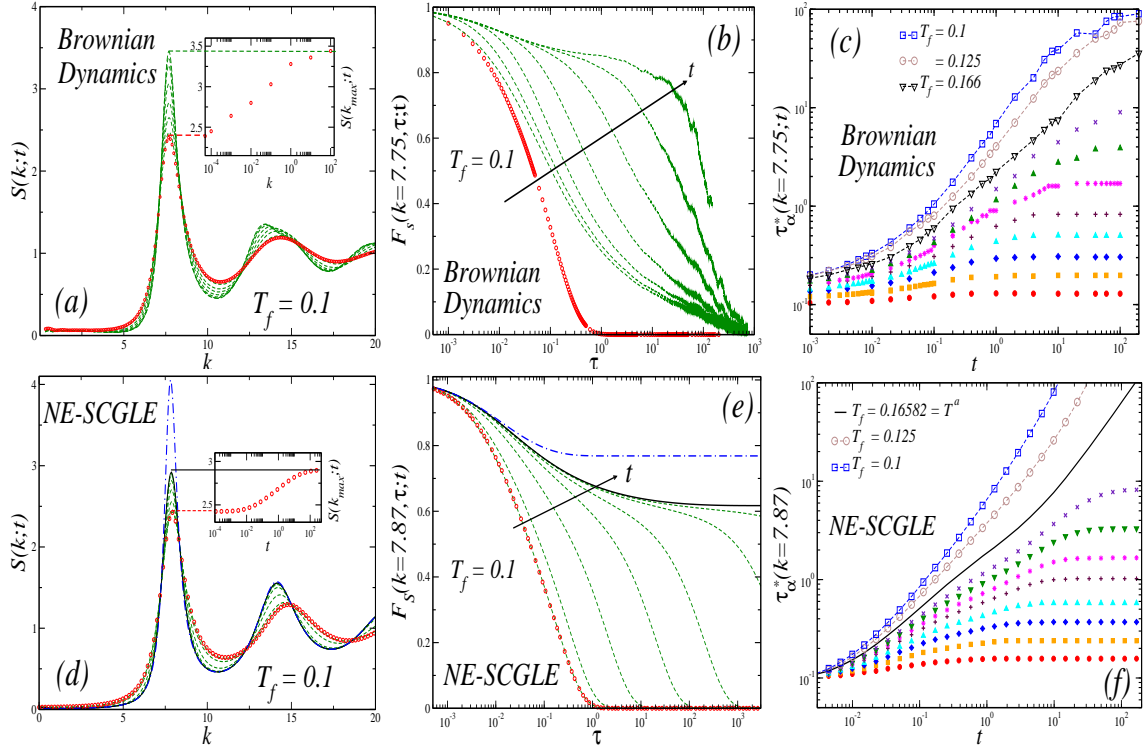

\includegraphics[width=0.3\textwidth, height=0.3\textwidth]{agrs/new_fig3a.eps}
\includegraphics[width=0.3\textwidth, height=0.3\textwidth]{agrs/new_fig3b.eps}
\includegraphics[width=0.3\textwidth, height=0.3\textwidth]{agrs/Final_Fig3C.eps}\\
\includegraphics[width=0.3\textwidth, height=0.3\textwidth]{agrs/Final_Fig3D.eps}
\includegraphics[width=0.3\textwidth, height=0.3\textwidth]{agrs/Final_Fig3E.eps}
\includegraphics[width=0.3\textwidth, height=0.3\textwidth]{agrs/Final_Fig3F.eps}
  \caption{Schematic representation of an aging process of the WCA 
  fluid after an instantaneous isochoric quench from the equilibrium
  state ($\phi=0.7,T_i=1$) towards a final temperature $T_f$ below 
  the dynamical arrest point $T^a=0.166$. (a) and (b) Sequence of 
  snapshots at waiting times $t=10^{-3},10^{-2},10^{-1},10^0,10^1,10^2$ 
  and $10^3$ (green dashed lines) obtained from BD simulations 
  describing the $t$-evolution of $S(k;t)$ and $F(k,\tau;t)$, 
  respectively, after an instantaneous quench to $T_f=0.1$. 
  In both figures, the initial ($t=0$) values are indicated in open 
  symbols. (c) Evolution of $\tau^*_{\alpha}(t)$ for three different 
  quenches below the dynamic arrest temperature as indicated (dashed 
  lines with open symbols). For reference, we show again the data of 
  Fig. \ref{fig:2}(c) (solid symbols).
  (d) and (e) Sequence of snapshots at waiting times 
  $t=10^{-3}, 10^{-2},10^{-1},10^0,10^1,10^2$ and $10^3$ (green 
  dashed lines) obtained with  the NE-SCGLE for $S(k;t)$ and 
  $F(k,\tau;t)$, respectively, for a quench with $T_f=0.1$. In these 
  figures, the solid lines represent the arrested values $S^a(k)$ and 
  $F^a_S(k,\tau)$, whilst the dashed-dotted lines correspond to 
  the thermodynamical input $[n\mathcal{E}^{(f)}(k)]^{-1}$ and the 
  prediction of the equilibrium SCGLE for $F^S(k,\tau:t\to\infty)$ 
  (see the text). 
  (f) Evolution of $\tau^*_\alpha(t)$ for a quench to the critical value 
  $T^a(\phi=0.7)=0.16582$ (solid line) and for two additional aging 
  processes (symbols with solid-lines) predicted by the NE-SCGLE for 
  quenches inside the region of dynamical arrest. For reference, we show 
  again the data of Fig. \ref{fig:2}(f) (solid symbols).}
  \label{fig:3}
\end{figure*}

As shown in Fig. \ref{fig:3}(a), for a deeper quench of the simulated 
system ($T_f=0.1$) the SF exhibits what at first sight seems essentially 
the same qualitative time-evolution as that found for the equilibration 
process described in Fig. \ref{fig:2}(a) (for $T_f=0.25$). As before, 
this evolution consists of a gradual increase of the structural 
correlations around the same wave vector number $k=k_{max}$ and the 
development of a (rather similar) stationary SF. 
A noteworthy difference with respect to the case of equilibration, 
however, is observed in the much slower relaxation of $S(k;t)$ 
towards a stationary value. One notices, for instance, that the 
time scale required to reach stationarity now extends over virtually 
the whole waiting time-window resolved in our BD simulations 
(see the inset of Fig. \ref{fig:3}(a)) and, also, that such time 
scale becomes roughly two orders of magnitude larger with respect 
to that described in the inset of Fig. \ref{fig:2}(a). Thus, upon a 
deeper quench, except for the much slower relaxation of the SF to 
acquire a new stationary value, the function $S(k;t)$ does not exhibit 
any other obvious qualitative difference with respect to an 
equilibration process.

This is to be contrasted with the dynamical behavior shown in Fig. 
\ref{fig:3}(b), which is fundamentally different from that of Fig. 
\ref{fig:2}(b) describing the equilibration of the dynamics. 
In this case, for example, one notices a more pronounced slowing 
down in the simulated system, highlighted by the gradual development 
of a plateau in $F(k;t)$ that leads to a much larger increase of 
$\tau_{\alpha}(t)$ at comparable waiting-times (up to two orders 
of magnitude). 
More crucially, for this quench the $\alpha$-relaxation time does 
not exhibit any evidence of stationarity within the waiting-time 
window of the simulations and, instead, it continues increasing 
with $t$. These features are reminiscent of an aging process in 
the dynamics \cite{kobbarrat,warren}, where the characteristic 
relaxation time (and, hence, the viscosity of the system) increases 
dramatically and persistently with system's age during the endless 
process of formation of a glassy state.

From the comparison of Figs. \ref{fig:2}(a) and \ref{fig:2}(b) with
Figs \ref{fig:3}(a) and \ref{fig:3}(b), respectively, one notices 
that two different quenches with rather similar stationary structural
correlations display an entirely different dynamical behavior, which 
is indicative of the crossover from the equilibriation to the glassy
regime. 
To emphasize such crossover, Fig. \ref{fig:3}(c) exhibits the 
simulation results for the aging of the $\alpha$-relaxation time 
corresponding to three representative quenches with final temperatures 
below the dynamical arrest temperature (open symbols with dashed lines). 
For reference, Fig. \ref{fig:3}(c) also reproduces the results for 
$\tau_{\alpha}(t)$ of the equilibration processes in Fig. \ref{fig:2}(c) 
(solid symbols), thus highlighting the transition from the previously 
discussed pattern of equilibration, where $\tau_{\alpha}(t)$ saturates 
to the value $\tau^{eq}_{\alpha}(T_f)$ after an equilibration time 
$t^{eq}(T_f)$, to a characteristic pattern of aging. In the latter,  
$\tau_{\alpha}(t)$ displays a notoriously larger and unbounded growth,  
with no equilibration time identifiable within the waiting time window 
of the simulations. Instead, what we observe is an increasingly longer
time interval in which $\tau_{\alpha}(t)$ grows as $\tau_{\alpha}(t) 
\propto t^\delta$, with a sub-aging exponent that gradually approaches 
the unity value as $T_f$ becomes smaller (see Sec. \ref{compa}). 
Beyond this interval, the aging effects on the dynamics become more 
sensitive on the observation time window, so that the evolution of the 
structural relaxation time is no longer described by a simple power 
law and follows, instead, and approximate functional relationship of 
the form $\tau_{\alpha}(t) \propto t^{\mu}$, with $\mu(t)=\mu[t;T_f]$ 
being the apparent instantaneous slope introduced in Sec. \ref{SEC2}.

\subsection{Aging of the structure of the WCA model: NE-SCGLE theory.}\label{agingSC1}

All the above mentioned non-equilibrium fingerprints observed in the 
kinetics of the structure and dynamics of the simulated system can be 
well understood within the framework of the NE-SCGLE. 
To see this, let us refer first to the most central equation of this 
theory, namely, Eq. \eqref{relsigmadif2pp} in appendix 
\ref{appendix_theo}, which describes the time-evolution of the SF 
after a quench towards a final state point $(n_f,T_f)$ and reads

\begin{equation}\label{Eq: dSdt}
\frac{\partial S(k;t)}{\partial t} = -2k^2 D^0
b(t)n_{f}\mathcal{E}^{(f)}(k) \left[S(k;t)
-1/n_{f}\mathcal{E}^{(f)}(k)\right]. \nonumber
\end{equation}

\noindent In this equation the mobility function $b(t)=D^L(t)/D^0$ is 
given in terms of the long-time self-diffusion coefficient $D^L(t)$ 
(at observation time $t$) and $\mathcal{E}^{(f)}(k)$ is the second
functional derivative of a postulated Helmholtz free energy
density-functional, evaluated at the uniform (bulk) density and
temperature fields $n(\mathbf{r},t)=n_f$ and $T(\mathbf{r},t)=T_f$.
As explained in detail in Refs. \cite{nescgle1,nescgle2,nescgle3},
this externally determined thermodynamical input plays a fundamental
role in the NE-SCGLE description of glassy states, with 
$[n_f\mathcal{E}^{(f)}(k)]^{-1}=S^{eq}(k)$ actually being the equilibrium 
structure factor at the state point ($n_f,T_f$). The function 
$n_f\mathcal{E}^{(f)}(k)$ can also be written as 
$n_f\mathcal{E}^{(f)}(k)= 1-n_fc[k;n_f,T_f]$, with $c[k;n_f,T_f]$ being 
the Fourier transform of the two-particle direct correlation function 
\cite{hansen}.

Thus, in connection with the previous discussion, let us consider the
stationary solutions ($\partial S(k,t)/\partial t=0$) of the above
equation, which might be classified in two different categories by mere
inspection. The first category are equilibrium solutions, occurring for
quenches with $T_f>T^a$ and determined by the condition
$\left[S(k;t)-1/n_{f}\mathcal{E}^{(f)}(k)\right]=0$, which implies that
$S(k;t)$ indeed attains its equilibrium expected value
$S^{eq}(k)=1/n_{f}\mathcal{E}^{(f)}(k)$ within a finite waiting-time
window.
The specific kinetic features of this kind of solutions were
already discussed and highlighted in the three lower panels of
Fig. \ref{fig:2}, finding a remarkable consistency with the results 
of BD. This includes the same essential evolution of the SF within
a comparable time-window, and the commensurable slowing down of the
microscopic dynamics.  More important, however, it also includes the 
rather similar power law predicted for the equilibration time-scale 
of the WCA model. Thus, equilibrium solutions are one of the possible 
stationary solutions of Eq. \eqref{relsigmadif2pp}.

A second category of stationary solutions, however, are those
describing dynamical arrest, corresponding to quenches inside the
non-ergodic region, or $T_f\leq T^a$.
In this case, the condition $\partial S(k;t)/\partial t=0$ is fulfilled
because the function $b(t)$ (that is, the diffusivity of the constituent
particles) decreases dramatically with $t$ (ideally  vanishing at 
$t=\infty$). This leads to a different kind of stationarity in the SF, 
that originates in underlying kinetic barriers and, hence, does not 
require the equilibrium condition
$\displaystyle{\lim_{t\to\infty}}S(k;t)=S^{eq}(k)$ to be satisfied.
This kind of solutions happens to correspond to glasses, gels, and other
non-equilibrium amorphous states, as it has been systematically 
illustrated in very specific circumstances in the context of 
glass-forming systems whose particles interact through repulsive 
\cite{nescgle3,gabriel,nescgle5}, attractive \cite{olais1,olais2,olais3} 
and even non-spherical potentials \cite{peredo1,cortes}.
For this scenario, the function $S(k;t)$ is predicted to evolve with
waiting-time towards a new stationary value $S_{a}(k)$ which, as 
revealed by the NE-SCGLE, is fundamentally different from the equilibrium 
structure factor $S^{eq}(k)$.

The importance of emphasizing this fundamental distinction can hardly 
be overstated. A noteworthy reason is that, as seen in the simulation 
results of Figs. \ref{fig:2}(a) and \ref{fig:3}(a), the difference 
between the stationary values attained by $S(k;t)$ for equilibration 
or dynamical arrest is, for all practical purposes, indistinguishable. 
This, for instance, might complicate -- or even impede -- to 
experimentally distinguish between equilibration and non-equilibrium 
aging at the level only of the static structural correlations 
represented by $S(k;t)$ (excepting for the much slower relaxation to 
reach a stationary value in the later case) 
Within the NE-SCGLE framework, in contrast, such distinction is 
explicitly recognized through the intrinsic difference between the 
predicted arrested value $S_{a}(k)$ (which, in general, depends on 
the protocol of fabrication) and the equilibrium structure factor 
$S^{eq}(k)$. It is worth mentioning here that such difference is 
generically enhanced as $T_f$ decreases \cite{peredo1} (always below 
the critical temperature $T^a$).

%
%

To provide a more vivid notion of these concepts, let us now refer to
Fig. \ref{fig:3}(d) which displays the arrested kinetics of the SF as 
described by the NE-SCGLE for a quench with 
$T_f=0.1 < T^a(\phi=0.7)=0.16582$.  
As observed before when comparing the simulation results of Figs. 
\ref{fig:3}(a) and \ref{fig:2}(a), the theoretically predicted results 
for $S(k;t)$ in Fig. \ref{fig:3}(d) also describe a qualitatively 
similar evolution compared to that obtained for the equilibration 
process shown in Fig. \ref{fig:2}(d). Also in agreement with our BD 
simulations, this relaxation takes place during a larger time-scale, 
after which $S(k;t)$ saturates to the stationary value $S_a(k)$. 
This non-equilibrium stationary SF, highlighted by the solid line in 
Fig. \ref{fig:3}(d), is notably different from the equilibrium structure 
factor $S^{eq}(k)= 1/n_f\mathcal{E}^{(f)}(k)$ expected at the final 
state, included for reference in the same figure (dashed-dotted line). 
Hence, according to our previous discussion, this indicates that the 
stationarity in $S(k;t)$ originates in the dynamical arrest conditions 
imposed by the long-time vanishing of the mobility $b(t)$.

\subsection{Aging of the dynamics of the WCA model: NE-SCGLE theory.}\label{agingSC2}

Besides the previously described non-equilibrium kinetics of the SF, 
the NE-SCGLE also provides predictions for the aging of the dynamics.
In Fig. \ref{fig:3}(e), for example, we show the theoretical results 
obtained for the $t$-evolution of the ISF after a quench with $T_f=0.1$.
Clearly, these results exhibit the same qualitative features as those 
reported by the simulations in Fig. \ref{fig:3}(b). For instance, one 
notices that the theory also predicts an aging process, characterized 
by the development of an increasingly long-lasting plateau in the 
function $F^S(k,\tau;t)$ and leading to a persistent growth of 
$\tau_{\alpha}(t)$ with $t$. 
In other words, the results of both simulations and theory support 
the interpretation that, at any \emph{finite} waiting-time $t$, 
the self ISF will always decay to zero with $\tau$ after a \emph{finite} 
$\alpha$-relaxation time $\tau_{\alpha}(t)$, in full agreement also with 
experimental data on glass forming liquids. 
This qualitative consistency between theory, simulations and experiments, 
however, is restricted by the practical limit of the longest possible
waiting times that are accessible in any simulation and/or experiment. 

The NE-SCGLE, however, still has meaningful predictions beyond such 
practical limitation, even if simulations or experiments cannot wait 
long enough to test them. For instance, the theory assures us that, at 
any finite waiting time $t$ beyond the aforementioned practical limit, 
the ISF will continue to decay to zero after a \emph{finite} 
$\alpha$-relaxation time $\tau_{\alpha}(t)$, regardless of the fact that 
we cannot measure such decay. 
It is only at the ideal asymptotic limit $t=\infty$, that $F_S(k,\tau;t)$ 
is predicted to never decay with $\tau$ (solid line in Fig. 
\ref{fig:3}(e)). Therefore, the theory itself advises that such 
idealized physical scenario is, for obvious reasons, unobservable in 
practice. 
This decoupling between the theoretical predictions of equilibrium 
approaches such as MCT and the SCGLE, and experimental or simulated 
observations, originates in the extremely slow kinetics of the aging 
processes that are characteristic of glass formation, as compared with 
the kinetics of equilibration, which always relaxes within finite 
equilibration times (except in the close vicinity of $T^a$). 

This realistic NE-SCGLE aging scenario thus complements and clarifies 
some fundamental aspects of the notion of the ``ideal glass transition'' 
predicted by both MCT and the SCGLE \cite{thvg_elizondo}. Recall that 
these theories only provide the equilibrium value $F_S^{eq}(k,\tau)$ of 
the ISF at any state point $(\phi,T)$, without any mention to the waiting 
time $t$. 
For reference, in Fig. \ref{fig:3}(e) we also show the prediction of 
the equilibrium SCGLE theory for $F_S^{eq}(k,\tau)$, calculated at the 
(arrested) state point $(\phi=0.7,T_f=0.1)$ (dash-dotted line). These 
equilibrium dynamical correlations involve an infinitely-lasting 
plateau, thus illustrating the MCT-like prediction that 
$\tau_{\alpha}^{eq}(\phi,T)$ is infinite (or equivalently, $D^L=0$) 
for $T\leq T^a(\phi)$. 
However, in the more general framework of the NE-SCGLE theory 
$F_S^{eq}(k,\tau)$ is only one possible asymptotic long-time limit 
of the function $F_S(k,\tau;t)$, \emph{i.e.,} 
$F_S^{eq}(k,\tau)=F_S(k,\tau;t\to\infty)$. In addition, for 
$T_f\leq T^a(\phi)$ another long-time asymptotic limit $F_S^{a}(k,\tau)$ 
exists (solid line in Fig. \ref{fig:3}(e)), representing non-equilibrium 
dynamically arrested states. Although both limits exist and compete, the 
spontaneous kinetic evolution after quenching favors the gradual 
development of the dynamically arrested limit $F_S^{a}(k,\tau)$. In 
contrast, for $T_f> T^a(\phi)$, neither $S_{a}(k)$ nor $F_S^{a}(k,\tau)$ 
exist, but $S^{eq}(k)$ and $F_S^{eq}(k,\tau)$ do, leaving the system 
with only one possibility, namely, to equilibrate.

Let us stress that there is a complete coincidence between the 
NE-SCGLE theory and its equilibrium (SCGLE) version, regarding the 
existence and location of the ``ideal'' glass transition line
$T= T^a(\phi)$ (solid line in Fig. \ref{fig:1}). As explained in Refs. 
\cite{nescgle1,nescgle2,nescgle3}, the predicted value of the critical 
temperature $T^a(\phi)$ turns out to coincide exactly with the critical 
temperature predicted by the equations of the equilibrium SCGLE theory. 
This is probably not that remarkable, since the NE-SCGLE theory is a 
natural non-equilibrium extension of this equilibrium theory (and, 
disregarding details, of MCT \cite{thvg_elizondo}). Thus, one might be 
tempted to extend the same criticism originally made to MCT, regarding 
the non-observability of this predicted ``ideal'' glass  transition 
line, to the NE-SCGLE theory. In the light of the present discussion, 
however, it is clear that this criticism doesn't apply in the present 
case, since the genuinely relevant comparison is between the real 
experimental measurements of $F_S(k,\tau;t)$ and $\tau_{\alpha}(t)$ 
at finite waiting times, and their theoretical counterparts, provided 
only by the NE-SCGLE theory.

This is illustrated here with the results in Fig. \ref{fig:3}(f), which 
display the behavior of $\tau_{\alpha}(t)$ as predicted by the NE-SCGLE 
for the same sequence of simulated instantaneous isochoric quenches in 
Fig. \ref{fig:3}(c). These quenches start from the same initial 
temperature $T_i$, and end at final temperatures $T_f$ falling above 
(solid symbols), at (solid line), and below (open symbols with dashed 
lines) the theoretically-determined dynamical arrest temperature 
$T^a(\phi=0.7)=0.166$. Just like the simulation results in Fig. 
\ref{fig:3}(c), these theoretical predictions also illustrate the 
passage from the characteristic pattern of equilibration 
($T_f>T^a$) to the dynamical arrest regime ($T_f\leq T^a$). 
Thus, according to the more realistic theoretical lens provided by the 
NE-SCGLE theory, the discontinuous dynamical arrest transition predicted 
by MCT and the equilibrium SCGLE theory actually appears, at any realistic 
finite waiting time $t$, as a broad crossover from full equilibration 
to full dynamic arrest. These are relevant issues that deserve to 
be discussed further, as we do in the Section \ref{SEC4}. Before doing 
so, however, we find useful to provide first a stringent comparison of 
theory and simulations.

\subsection{Comparison of simulations and theory}\label{compa}

\begin{figure*}
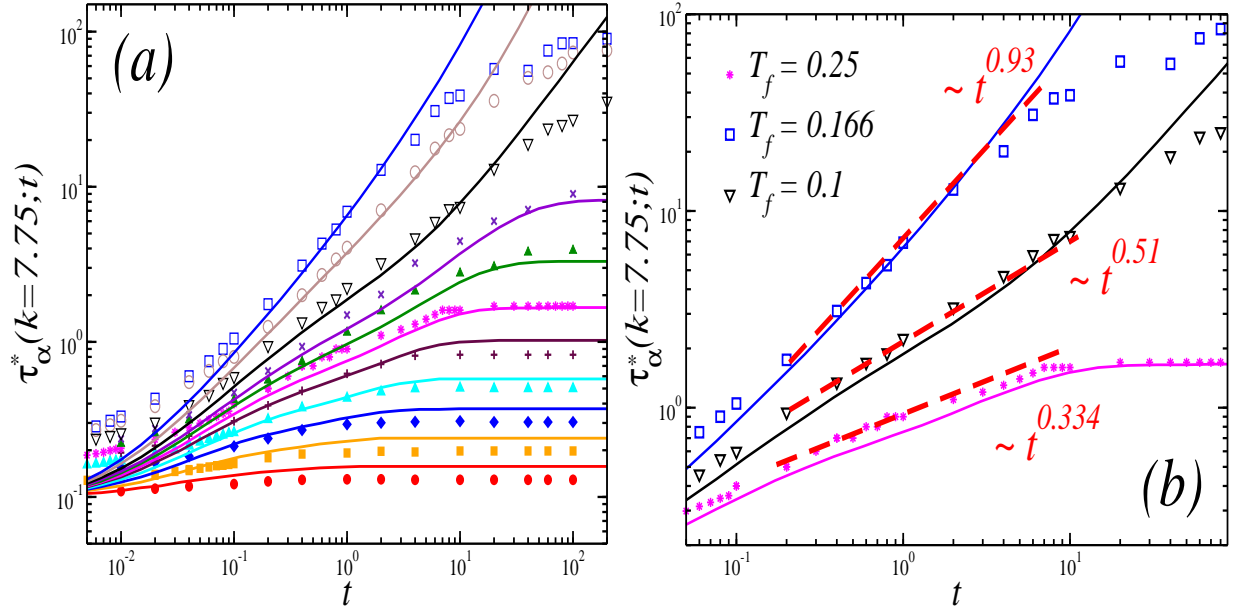

\includegraphics[width=0.49\textwidth, height=0.49\textwidth]{agrs/Fig_comp_a.eps}
\includegraphics[width=0.49\textwidth, height=0.49\textwidth]{agrs/Fig_comp_b.eps}
  \caption{(a) Comparison of the results in Figs. \ref{fig:3}(c) and 
  \ref{fig:3}(f) for $\tau^*_{\alpha}(k;T_f;t)$, as a function 
  of waiting time $t$. Symbols correspond to simulations results, 
  with the same nomenclature used in Fig. \ref{fig:3} for the 
  $T_f$-values of the sequence of quenches, whereas lines describe 
  the corresponding results of the NE-SCGLE shown in Fig. \ref{fig:3}(f).   
  Same information as in (a), but only for the three representative 
  quenches above ($T_f=0.25$), at ($T_f=T^a=0.166$) and below 
  ($T_f=0.1$) the ideal glass transition temperature $T^a=0.166$. 
  The thick dashed lines describe power laws of the form 
  $\tau^{\alpha}\sim t^{\delta}$, with exponent (from bottom to top) 
  $\delta=0.334,0.51,0.93$, obtained from a fit of the simulation 
  data for $t\in[10^{-1},10^1]$ (see the text). 
  }
  \label{fig:4}
\end{figure*}

The results in Figs. \ref{fig:3}(c) and \ref{fig:3}(f) are highly 
instructive and revealing since, combined, they provide a 
consistent kinetic description of the isochoric processes of 
equilibration and aging occurring in the WCA liquid. To better 
compare theory and simulations at realistic waiting times, in 
Fig. \ref{fig:4}(a) we plot together such results. There we can 
observe the remarkable quantitative agreement at all waiting 
times in the equilibration regime ($T_f>T^a=0.166$).
For the three quenches with $T_f \le T^a$ (i.e., in the dynamic 
arrest regime), theory and simulations also coincide in predicting a
stronger slowing down of the dynamics, represented by an increasingly 
faster development of $\tau_{\alpha}(t)$. 

In both, the equilibration and dynamical arrest scenarios, one also
observes an excellent semi-quantitative agreement at intermediate 
waiting times. As mentioned above, in this $t$-regime $\tau_{\alpha}(t)$ 
grows as a power law of $t$, $\tau_{\alpha}(t) \propto t^\delta$, 
with a $T_f$-dependent sub-aging exponent $\delta<1$. 
The emergence of this sub-aging regime is better illustrated in 
Fig. \ref{fig:4}(b), which considers three representative quenches 
above ($T=0.25$), at ($T=0.166$) and below ($T_f=0.1$) the critical 
value $T^a$.  
For reference (and to serve as a guide for the eye) in the same 
figure we also plot three different power laws in thick dashed 
lines, obtained from a fit of the simulated data for 
$t\in[10^{-1},10^1]$ and which highlight the remarkable coincidence 
of theory and simulations at intermediate $t$. Notice that, in both 
approaches, $\delta\to1$ as $T_f$ becomes smaller.

For quenches with $T_f\leq T^a$, however, the  agreement of theory
and simulations visibly deteriorates at longer $t$, with the theoretical 
$\tau_{\alpha}(t)$ growing even faster, as an asymptotic power law 
$\tau_{\alpha}(t)\propto t^\gamma$, with $\gamma > 1$ (hyper-aging) 
for the two quenches with $T_f< T^a$, and with $\gamma = 1$ (normal 
aging) for the quench to the critical temperature $T^a=0.166$ 
\cite{rizzo3}. As seen in Fig. \ref{fig:4}(b), the simulation results 
strongly deviate from this predicted asymptotic behavior.  
The origin of these deviations, however, is not difficult to understand. 
Although there may be a variety of reasons, the most relevant ones we 
can think of, are the deviations $\Delta \overline{n}(\mathbf{r};t) 
\equiv \overline{n}(\mathbf{r};t)-n$ of the local mean particle density
$\overline{n}(\mathbf{r};t)$ from its bulk value $n$. These deviations 
describe the spatial heterogeneity of the local density and, 
consequently, the emergence of dynamical heterogeneities, 
unavoidably present in microscopic simulations and in real experiments. 

The original and most 
general version of the NE-SCGLE theory, summarized by Eqs. (4.1)-(4.7) 
of Ref. \cite{nescgle1}, was posed precisely  in terms  of the  
time-evolution equations for the mean value $\overline{n}(\textbf{r},t)$ 
and covariance $\sigma(\textbf{r},\textbf{r}';t)$, of the instantaneous 
local particle density $n(\textbf{r},t)$ of the fluid (see Eqs. 
(4.1)-(4.3) of Ref. \cite{nescgle1}). The solution of such time-evolution 
equations, which are coupled between them through a state-dependent local 
mobility function $b(\textbf{r},t)$,  in principle include the description 
of these spatial (structural and dynamical) heterogeneities. 
Unfortunately,  $b(\textbf{r},t)$ must be determined by a self-consistent 
system of equations (Eqs. (4.4)-(4.7) of Ref. \cite{nescgle1}), which 
involve spatially-varying non-equilibrium dynamical properties. The 
numerical solution of the resulting full system of equations poses 
a formidable technical challenge.    
  
To have a glimpse of the physical scenario revealed by this general 
non-equilibrium theory, however, a provisional simplification was  
introduced  \cite{nescgle1}, in which one solves these equations 
ignoring precisely these spatial heterogeneities. For this, we neglect 
the deviations $\Delta \overline{n}(\mathbf{r};t)$, in which case 
$\overline{n}(\mathbf{r};t)=n$, and hence, we do not need the time 
evolution equation for $\overline{n}(\mathbf{r};t)$ (Eq. (4.1) of Ref. 
\cite{nescgle1}). Within this approximation,  Eqs. (4.3)-(4.7)  of Ref. 
\cite{nescgle1} become Eqs. (\ref{relsigmadif2pp})-(\ref{fluctsquench}) 
of Appendix  \ref{appendix_theo}, which are precisely the equations 
representing  the simpler version of the NE-SCGLE theory employed in 
the present study. Clearly, to go beyond this level of approximation 
we would need some strategy to incorporate the deviations 
$\Delta \overline{n}(\mathbf{r};t)$. 
The resulting $\overline{\tau}_{\alpha}(t)$ is expected to agree with 
the simulations in exhibiting similar deviations from the simplified 
theoretical behavior described by the solid lines of Fig. \ref{fig:4}(a). 

Implementing this or other strategy to incorporate spatial heterogeneities 
in the NE-SCGLE description is, however, not the main focus of the present 
paper, which is aimed, instead, at exhibiting how  accurately the 
simpler current version of the NE-SCGLE theory agrees with the simulated aging of the system in the realistic  regime of short and intermediate times.  In the meanwhile, 
however, let us mention that the generic behavior of the simulation 
results in Fig. \ref{fig:4} regarding their long-time deviations from theory, can also be understood from a rigorous 
mathematical analysis of the solutions of the NE-SCGLE equations 
following, to a large extent, the steps of Ref. \cite{thvg_elizondo} 
for the asymptotic solutions of the equilibrium SCGLE theory. Such an 
analysis allows to identify the \emph{closeness} of the structural 
relaxation for ``arbitrary'' $T_f$ in terms of an expansion parameter 
$\sigma=|1-(T_f/T^a)|$ (with $\sigma << 1$). The resulting theoretical 
scenario includes the prediction of the asymptotic normal and 
hyper-aging regimes of the dynamics for, respectively, $T_f=T^a$ 
and $T_f< T^a$. Although in practice these asymptotic regimes are 
interrupted by the spatial heterogeneities, the emerging dynamical 
heterogeneities can also be taken into account by merging the 
NE-SCGLE approach with the recently developed stochastic 
$\beta$-relaxation theory \cite{rizzo1,rizzo2}, with the former 
allowing to include non-mean-field fluctuations which account for 
the aforementioned deviations in the long $t$ regime. A thorough 
discussion of these aspects can be consulted in Ref. \cite{rizzo3}.

\section{Universal crossover from equilibration to  aging.}\label{SEC4}


Let us now focus on our main objective of emphasizing the 
strengths of the current version of the theory, even without including 
heterogeneities. In this sense, perhaps the farthest-reaching prediction 
of the simplified NE-SCGLE equations 
\eqref{relsigmadif2pp}-\eqref{lambdadk}, is the existence of the 
critical temperature $T^a(\phi)$ that separates the two possible 
(and mutually exclusive) physical scenarios of equilibration 
(when $T_f>T^a(\phi)$) and aging (when $T_f<T^a(\phi)$). 
The existence of this critical temperature $T^a(\phi)$ was first predicted by MCT (and later also by the equilibrium SCGLE theory), but only referring to the long-time asymptotic regime. The NE-SCGLE theory extends this prediction to any finite time after the quench, thus providing a  
richer and more explicit conceptual scenario, that can be employed 
to understand the phenomenology of the aging of real and simulated 
materials, severely constrained by the finite time-span of practical 
observations. In fact, the solution of these equations yields the time-evolution of the most 
relevant properties ($S(k;t)$, $b(t)$, $F_S(k;\tau;t)$, $\tau^*_{\alpha}
(t,T_f)$, etc.)  at \emph{any finite} waiting times, starting from the experimentally accessible short and intermediate times, to the often inaccessible asymptotically long  times.

\begin{figure*}[h!]
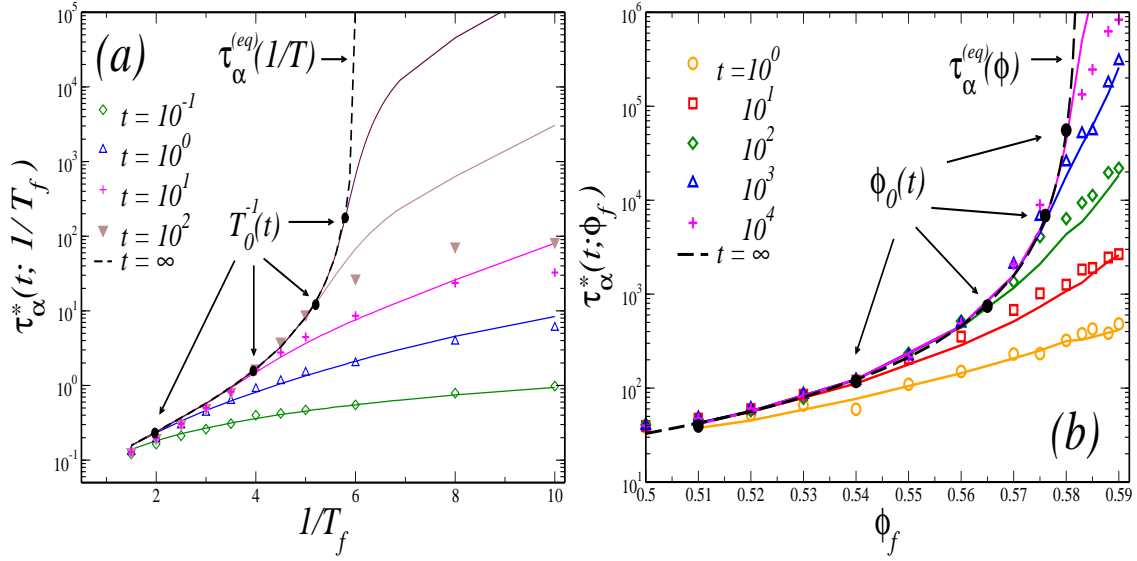
 
 \includegraphics[width=0.45\textwidth, height=0.45\textwidth]{agrs/New_Fig5A.eps}
 \includegraphics[width=0.45\textwidth, height=0.45\textwidth]{agrs/New_Fig5B.eps}
\caption{(a) Same information of Figs. \ref{fig:3}(c) and \ref{fig:3}(f) 
for the $\alpha$-relaxation time $\tau^*_{\alpha}(k;T_f;t)$, but now 
displayed as a function of the inverse temperature $1/T_f$ at fixed 
waiting time $t$, as indicated. Symbols are used for the simulation data 
whereas lines correspond to the results of the NE-SCGLE.
The black solid circles highlight the crossover value $1/T_0(t)$ that 
separates samples which have equilibrated from those which are 
insufficiently 
equilibrated, for distinct waiting times (see the text). The dashed 
line corresponds to the prediction of the equilibrium SCGLE for 
$\tau_{\alpha}^{(eq)}(k;1/T)$. 
(b) Molecular dynamics simulation data (symbols) and theoretical results 
(solid lines) reported in Ref. \cite{nescgle5} for the volume fraction 
dependence of $\tau^*_{\alpha}(k;\phi_f;t)$ at different waiting times, 
as indicated, for quenches of the WCA model to a final temperature 
$T_f=0$ and approaching the critical value $\phi_{HS}^a$. As in (a), 
the dashed line corresponds to the prediction of the SCGLE for 
$\tau_{\alpha}^{(eq)}(k;\phi)$.}
\label{fig:5}
\end{figure*}

In order to illustrate these concepts more clearly, let us now
present the same information in Fig. \ref{fig:4}(a), but now in a 
complementary format, by plotting  $\tau^*_{\alpha}(t;1/T_f)$ as a 
function of the inverse temperature $1/T_f$ for a sequence of fixed 
waiting times. 
Fig. \ref{fig:5}(a) displays such information, which shows that 
simulations and theory agree in reporting two distinct regimes as 
a function of $1/T_f$ in these plots. 
To see this, notice that each theoretical curve for 
$\tau^*_{\alpha}(t;1/T_f)$ at fixed waiting time $t$, exhibits a
regime corresponding to samples that, at time $t$, have fully 
equilibrated ($1/T_f\leq 1/T_0(t)$), followed by a second regime, 
corresponding to samples that have not yet reached equilibrium 
($1/T_f\geq 1/T_0(t)$). 
These two regimes are separated by a crossover inverse temperature 
$1/T_0(t)$, highlighted by the black solid circles in the figure. 
Focusing on the simulation results for $t=10^1$, for instance, we 
find a crossover value $1/T_0(t=10^1)\approx 4.13$ 
(i.e., $T_0(t=10^1)\approx0.242$), 
whereas for $t=10^2$ we find $1/T_0(t=10^2)\approx 5.21$ 
($T_0(t=10^2)\approx0.192$). 
One notices that the crossover  $[1/T_0(t)]$ found in simulations first increases 
rather fast with $t$ and then tends to saturate, with the crossover points following closely the ideal 
curve $\tau_{\alpha}^{(eq)}(1/T)$ (dashed line) predicted by the 
equilibrium SCGLE theory, which diverges as $T$ approaches  the critical value 
$T^a=0.16582$. The theoretical prediction is that  $[1/T_0(t)]\to [1/T^a]$ from below (i.e., 
$T_0(t)\to T^a$ from above) as $t\to\infty$, which is the trend confirmed through the duration ($t\leq t_{max}\approx 10^2$) of our simulations. The results in Fig. \ref{fig:4}(a) then illustrate the origin of the un-observability of any divergence of $\tau^*_{\alpha}(t)$ in our BD simulations, particularly for quenches with $T_f$ well below the critical temperature $T^a$. In fact, the theoretical prediction is that the same scenario will prevail no matter how long the simulations (or experiments!) last, since they will always involve finite waiting times.

In fact, essentially the same physical scenario was found in Refs. 
\cite{gabriel,nescgle5} in considerably longer molecular dynamics 
simulations of a sequence of quenches of the WCA model to the final 
temperature  $T_f=0$, but along a sequence of isochores with increasing 
volume fraction $\phi$. This was used to investigate the density-driven 
hard-sphere glass transition of the HS model by increasing the volume 
fraction $\phi_{HS}$ towards its critical value $\phi^a_{HS}=0.582$. To 
serve as a reference, in Fig. \ref{fig:4}(b) 
we reproduce the plot of $\tau^*_{\alpha}(t;\phi_f)$ vs $\phi_{HS}$ 
at different waiting times, extracted from Fig. 3 of Ref. \cite{nescgle5}, 
which compared molecular dynamics (MD) simulations and theory. 
Just as  in Fig. \ref{fig:4}(a), one observes that for quenches with 
$\phi_{HS}\geq0.582$ the $t$-dependent $\alpha$-relaxation time does 
not saturate to its equilibrium value within the total duration of 
the MD simulation runs. 
Therefore, in both cases (Figs. \ref{fig:4}(a) and \ref{fig:4}(b))  the 
comparison between the results of simulations of duration $t_{max}$ and the 
predictions of equilibrium theories, such as  SCGLE or MCT, is only 
meaningful for samples below the crossover thresholds $1/T_0(t_{max})$ and  
$\phi_0(t_{max})$, respectively.

Let us also emphasize that the noticeable qualitative similitude of the 
overall physical scenario outlined by Figs. \ref{fig:4}(a) and 
\ref{fig:4}(b), along with all the results presented so far, provides a 
vivid confirmation of the intuitive notion that the inverse temperature 
$1/T$ plays an analogous role to that of the volume fraction 
$\phi$ in determining the structural relaxation of the WCA model near 
the glass transition. We should us stress here that this physical notion is 
also in agreement with experimental data for soft-sphere colloidal 
suspensions approaching the GT upon variations in both parameters 
\cite{rivas}. 
In addition, let us mention that a full correspondence between the two 
control parameters may also be established more explicitly when extending 
to non-equilibrium conditions the principle of dynamic equivalence between 
soft- and hard-sphere fluids \cite{ramirez2011}. Such discussion, however, 
is out of the scope of the present contribution and is left for further 
work.

\section{Concluding remarks}\label{conclusions}

In summary, in this work we have combined Brownian dynamics simulations 
and NE-SCGLE theoretical calculations to present a consistent physical 
description of the irreversible isochoric relaxation of a soft-sphere 
glass-forming liquid suddenly quenched towards conditions of dynamical 
arrest.
Such relaxation was analyzed in terms of the structural and dynamical
properties of the model liquid, whose waiting time evolution falls in 
two mutually exclusive possibilities:
either the system equilibrates within a finite equilibration time-scale
$t^{eq}(T_f)$ that depends on the final temperature of the
quench, or the system becomes dynamically arrested and ages endlessly
during the process of becoming a glass. 

The distinction between these two scenarios was based on the theoretical predictions of the dynamical behavior of the liquid after quenching.
The evolution of the $\alpha$-relaxation time $\tau_{\alpha}(t)$, in
particular, allowed for the description of the crossover from the
equilibration to the aging regimes, and to highlight various
non-equilibrium fingerprints that characterize the relaxation of the
model system studied. In the case of equilibration (that is, for 
quenches above the critical temperature, $T_f>T^a$), both, 
$\tau_{\alpha}^{eq}(T_f)$ and the equilibration time $t^{eq}(T_f)$ 
required to reach thermodynamic equilibrium, are found to remain finite, 
with $t^{eq}(T_f)$ growing as a power law of $\tau_{\alpha}^{eq}(T_f)$ 
and with both quantities showing a significant increase as $T_f$ 
approaches $T^a$ from above. For quenches to $T^a$ or below, instead, $\tau_{\alpha}(t)$ undergoes a 
persistent growth with system's age within the whole 
waiting-time resolution employed, showing also 
an increasingly faster growth for smaller $T_f$.
These dynamic features were consistently found in our simulations and 
in the predictions of the non-equilibrium self-consistent generalized 
Langevin equation theory.


At the level of structural correlations, in contrast, the two relaxation
scenarios turn out to be very similar and, therefore, too difficult (if
not impossible) to distinguish in the simulations.
For both, equilibration and dynamical arrest, the static structure factor
evolves towards a stationary value which does not exhibit any noticeable  qualitative
 difference.
The theoretical framework employed here, however, has shown to provide a
powerful microscopically founded tool for the understanding of the two
relaxation processes at the level of both, the dynamical and the structural 
properties.
In the latter case, for instance, the theory explicitly recognizes and 
emphasizes the fundamental differences between equilibration and dynamical 
arrest in terms of the predicted long time asymptotic limit of $S(k;t)$  
which, for equilibration, reaches its thermodynamic equilibrium value 
$S^{eq}(k)$, while for dynamical arrest attains a non-equilibrium (and 
theoretically well defined) arrested value $S^a(k)$.
As explained in Sec. \ref{agingSC1}, despite the misleading quantitative
similarity between $S^{eq}(k)$ and $S^a(k)$, the intrinsic difference
between these stationary values plays a fundamental role in determining 
the dynamical behavior of the system. The theory highlights such difference
from the very beginning through its most central equation, which relates
the mobility of the constitutive particles with the stationary values 
through a highly non-linear functional dependence.

In our present contribution we have demonstrated the qualitative 
equivalence between the packing fraction and the inverse temperature 
as control parameters determining the non-equilibrium relaxation of a 
soft-sphere liquid. Furthermore, our results provide convincing evidence 
of the universality of the aging exponents accounting for the slowing 
down of the dynamics in systems described by short-range stiff potentials, 
as compared to independent simulation results \cite{warren}. 

Despite the various approximations adopted at the level of structure 
and dynamics for the theoretical description of the WCA liquid, the 
qualitative and semi-quantitative agreement between theory and 
simulations provides a rigorous validation of the non-equilibrium 
self-consistent generalized Langevin equation approach as a fundamental 
tool for the understanding of dynamical arrest phenomena in thermally 
driven glass-forming liquids.
Thus, we expect that the methodology and analysis presented in this work
could serve as a benchmark for the study and characterization of glass
formation in more complex systems involving, for instance, mixtures 
of thermo sensitive particles, fluids involving anisotropic (e.g., dipole-dipole) interactions 
\cite{cortes,peredo1}, and even more complex conditions.

\vskip2cm

\textbf{ACKNOWLEDGMENTS}: 
LFEA and FPV acknowledge the Consejo Nacional de Ciencia y Tecnolog\'{\i}a (CONACYT, M\'exico) for support through a Postdoctoral Fellowship (Grant No. I1200/224/2021). MMN, RPO, ELL, MCP and HRE are also grateful to CONACYT for financial support through grants No. 320983, CB A1-S-22362, 
No. CF-2019-731759, and LANIMFE 314881.

\vskip0.5cm

\textbf{DATA AVAILABILITY}:
The data that support the findings of this study are available from the corresponding author upon reasonable request.

\newpage

\appendix

\section{Theoretical considerations.}\label{appendix_theo}

In this appendix we summarize the essence of the simplified 
NE-SCGLE theory used to investigate the relaxation of the WCA 
model system.

\subsection{Review of the NE-SCGLE theory.}\label{review}

The NE-SCGLE theory was first derived in Ref. \cite{nescgle1} and
is summarized by a set of coupled time-evolution equations, whose
solution (for $t>0$) describes the irreversible relaxation of an
instantaneously quenched liquid. The most central of these equations
describes the waiting-time evolution of the static structure factor
$S(k,t)$. For an homogeneous system, instantaneously quenched at
$t=0$ from an initial equilibrium state $(n_i,T_i)$ and towards a
final state $(n_f,T_f)$, such equation reads

\begin{equation}
\frac{\partial S(k;t)}{\partial t} = -2k^2 D^0
b(t)n_{f}\mathcal{E}(k;n_f,T_f) \left[S(k;t)
-1/n_{f}\mathcal{E}(k;n_f,T_f)\right].
\label{relsigmadif2pp}
\end{equation}

\noindent In this equation, $D^0$ is the short-time self-diffusion
coefficient, related by Einstein's relation with the corresponding
short-time friction coefficient $\zeta^0=k_BT/D^0$ which, for colloidal
liquids, can be determined by its Stokes expression. The function
$\mathcal{E}(k;n_f,T_f)$, on the other hand,
is the Fourier transform (FT) of the functional derivative
$\mathcal{E}[\mid\textbf{r}-\textbf{r}'\mid;n,T] \equiv \left[
{\delta \beta\mu [{\bf r};n]}/{\delta n({\bf r}')}\right]$ of the
chemical potential $\mu [{\bf r};n]$ with respect to the local
concentration profile $n({\bf r})$, evaluated at the uniform concentration
and temperature profiles $n({\bf r})=n_f$ and $T({\bf r})=T_f$ of the final
state. This property is a fundamental thermodynamic input of the theory, whose practical 
determination is discussed below.

The time-evolving mobility function $b(t)$ appearing in the right side
of Eq. \eqref{relsigmadif2pp} is defined as $b(t)\equiv D_L(t)/D^0$,
with $D_L(t)$ being the long-time self-diffusion coefficient of the colloidal
particles at evolution time $t$. As explained in Refs. \cite{nescgle1,nescgle3},
the equation

\begin{equation}
b(t)= [1+\int_0^{\infty}
d\tau\Delta{\zeta}^*(\tau; t)]^{-1}
\label{bdt}
\end{equation}

\noindent relates $b(t)$ with the $t$-evolving and $\tau$-dependent friction
coefficient $\Delta{\zeta}^*(\tau; t)$, given approximately by

\begin{equation}
\begin{split}
  \Delta \zeta^* (\tau; t)= \frac{D_0}{24 \pi
^{3}n_f}
 \int d {\bf k}\ k^2 \left[\frac{ S(k;
t)-1}{S(k; t)}\right]^2  F(k,\tau; t)F_S(k,\tau; t).
\end{split}
\label{dzdtquench}
\end{equation}
Thus, the presence of $b(t)$ in Eq. (\ref{relsigmadif2pp}) couples $S(k;t)$
with the non-stationary two-time density correlation functions
$F(k,\tau; t)\equiv \displaystyle{\langle \sum_{n,n'}^N \exp(i\mathbf{k}\cdot[
\mathbf{r}_n(t+\tau)-\mathbf{r}_{n'}(t)]) \rangle}$ and $F_S(k,\tau; t)=
\langle \exp[i\mathbf{k}\cdot \Delta\mathbf{r}_T(\tau,t)]\rangle$,
for which the NE-SCGLE also provides time evolution equations. In terms
of their Laplace transforms (LT) $F(k,z; t)$ and $F_S(k,z; t)$, such
equations read,

\begin{gather}\label{fluctquench}
 F(k,z; t) = \frac{S(k; t)}{z+\displaystyle{\frac{k^2D^0 S^{-1}(k;
t)}{1+\lambda (k)\ \Delta \zeta^*(z; t)}}},
\end{gather}

and

\begin{gather}\label{fluctsquench}
 F_S(k,z; t) = \frac{1}{z+\displaystyle{\frac{k^2D^0 }{1+\lambda (k)\ \Delta
\zeta^*(z; t)}}},
\end{gather}

\noindent with $\lambda (k)$ being a phenomenological interpolating
function \cite{gabriel}, given by

\begin{equation}
\lambda (k)=1/[1+( k/k_{c})
^{2}],
\label{lambdadk}
\end{equation}
with $k_c$ being an empirically-chosen cutoff wave-vector. This parameter can be employed to calibrate the theory in each specific application, as indicated in Section \ref{SEC2}.

The solution of Eqs. (\ref{relsigmadif2pp})-(\ref{fluctsquench}) provides the t-evolution of the functions $S(k;t)$, $b(t)$, $\Delta\zeta^*(t)$, $F(k,\tau;t)$ and $F_S(k,\tau;t)$, which describe the irreversible relaxation
of an instantaneously and homogeneously quenched liquid. Specific details regarding the numerical solution of
Eqs. (\ref{relsigmadif2pp})-(\ref{fluctsquench}) can be found in
Ref. \cite{nescgle3} and access to a computational
code to solve these equations is provided in \cite{zepeda}.

\section{Simulation details}\label{sim_details}

Our Brownian dynamics simulations of $N$ particles interacting
through the soft repulsive potential in Eq. \eqref{truncatedlj}, 
were conducted in a cubic simulation box of volume $V$ with 
periodic boundary conditions. The positions of the particles in 
the simulation cell were updated according to the conventional 
Ermak-McCammon algorithm \cite{allen}

\begin{equation}
{\bf r}(t + \Delta t) = {\bf r}(t) + \beta D^0 {\bf F}(t) + \delta{\bf R},
\end{equation}

\noindent
where $\Delta t$ is the time step, ${\bf r}(t)$ and ${\bf F}(t)$ the 
position and force on the particle at time $t$, and $\delta {\bf R}$ 
represents, for each Cartesian component, a random displacement 
extracted from a Gaussian distribution with zero mean and variance 
$2D^0 \Delta t$.  
Our simulations followed the quench protocol introduced in this work. 
The samples were prepared and lead to equilibrium at the temperature 
$T_i$, and then quenched to the target temperature $T_f$ using 
$N=10000$ particles. 
This stage of the simulations was conducted for at least $500t_B$ 
(where $t_B=\sigma^2 /D^0$ is the Brownian time), using a step time 
$\Delta t=10^{-4}t_B$ for most of the systems; for some systems 
$\Delta t=5\times 10^{-5}t_B$ was used.
To simulate polydisperse systems we introduced a size polydispersity 
$P$ by taking the diameters of the $N$ particles evenly distributed 
between $\bar\sigma(1-w/2)$ and $\bar\sigma(1+w/2)$, with 
$\bar\sigma$ the mean diameter of the distribution and for
the particular value of $P=0.1$. Further details of our simulations 
can be found in Refs. \cite{gabriel} and \cite{lopez1}.

\end{document}